\documentclass{article}

\usepackage{amstext}
\usepackage{amssymb}
\usepackage{graphicx}
\usepackage{comment}
\usepackage{geometry}
\usepackage{hyperref}
\usepackage{verbatim}

\usepackage{url} 
\urldef{\mailsa}\path|{alfred.hofmann, brigitte.apfel, ursula.barth, christine.guenther,|
\urldef{\mailsb}\path|ingrid.haas, frank.holzwarth, anna.kramer, leonie.kunz, nicole.sator,| 
\urldef{\mailsc}\path|erika.siebert-cole, peter.strasser, lncs}@springer.com| 

\newtheorem{lemma}{Lemma}
\newtheorem{theorem}{Theorem}
\newtheorem{claim}{Claim}

\newcommand{\ptrees}{phylogenetic trees}

\begin{document}


\title{A 3-factor approximation algorithm for a Minimum Acyclic Agreement Forest on $k$
rooted, binary phylogenetic trees}

\author{Asish Mukhopadhyay \thanks{Research supported by an NSERC Grant to this author} \\ 
{and} Puspal Bhabak \thanks{Now a doctoral student at Concordia University, Montreal}  \\
School of Computer Science, University of Windsor, Windsor, ON N9B3P4 \\ 
E-mail: \tt{mailtobumba@gmail.com}, \tt{asishm@uwindsor.ca}}

\maketitle

\begin{abstract}

Molecular phylogenetics is a well-established field of research in biology wherein phylogenetic trees are 
analyzed to obtain insights into the evolutionary histories of organisms.
Phylogenetic trees are leaf-labelled trees, where the leaves correspond to extant species (taxa), and the 
internal vertices represent ancestral species.  
The evolutionary history of a set of species can be explained by more than one phylogenetic tree, 
giving rise to the problem of comparing phylogenetic trees for similarity. Various distance metrics, like the subtree prune-and-regraft (SPR),
tree bisection reconnection (TBR) and nearest neighbour interchange (NNI) have been proposed to capture this similarity.    
The distance between two phylogenetic trees can also be measured by the
size of a Maximum Agreement Forest (MAF) on these trees, as it has been shown that the rooted subtree prune-and-regraft distance
is 1 less than the size of a MAF. Thus, the smaller this size, the greater is the similarity between the two trees.  
Since computing a MAF of minimum size is an NP-hard problem, researchers have turned their attention 
to computing MAFs that approximate the minimum.    
Recently, it has been shown that the MAF on $k (\geq 2)$ trees can be approximated to within a factor of 8. 
In this paper, we improve this ratio to 3. For certain species, however, the evolutionary history is not completely
tree-like. Reticulation events, such as horizontal gene transfer (HGT),
hybridization and recombination have played a significant role in the evolution of these species. 
Suppose we have two phylogenetic trees each of which is for a gene of the same set of species. 
Due to reticulate evolution the two gene trees, though related, appear different, making
a phylogenetic network a more appropriate representation of 
reticulate evolution. A phylogenetic network contains hybrid nodes for the species evolved from two parents. 
The number of such  nodes is its hybridization number. It has been shown that this 
number is 1 less than the size of a Maximum Acyclic Agreement Forest (MAAF).  
We show that the MAAF for $k (\geq 2)$ phylogenetic trees can be approximated to within a factor of 3. 
       
\end{abstract}

\section{Introduction}

Phylogenetic trees, or evolutionary trees, are used in evolutionary
biology to represent the evolutionary history of an extant set of species. 
In a rooted phylogenetic tree, the leaves are uniquely labelled by these species, while the unlabelled
internal nodes represent their ancestors. The root represents  
the universal common ancestor of all the species. \\  
 
In a phylogenetic tree, the \emph{out-degree} of an internal node is the number
of its children. The distance between two nodes 
represents an evolutionary distance such as time or the number of mutations.
This kind of representation is appropriate for many groups of species
which include the mammals. \\  

It has been observed the evolutionary patterns are not the same for all groups.  
Sometimes, reticulation events come into play that do not conform to a 
tree-like evolutionary process. Rather, the species under reticulation
events form a composite of genes derived from different ancestors. Indeed, reticulation
involves a gamut of events that includes hybridization, horizontal gene transfer and recombination.
In this paper, we focus primarily on hybridization. \\  

Research over the years into the evolutionary history of Eukaryotes has revealed the existence
of hybridization events among certain groups of plants, birds
and fish. Spontaneous hybridization events have also been reported in
the evolutionary history of some mammals and even primates. Study
on hybridization shows that at least 25\% of plant species and 10\%
of animal species, mostly the youngest species, are involved in hybridization
events \cite{key-10}. \\ 

Several techniques have been devised to reconstruct phylogenetic trees
for a given set of species. This got Computational Biologists interested in the 
problem of determining the 'distance' between two such trees. Distance metrics such as NNI
(Nearest Neighbor Interchange), SPR (Subtree Prune and Regraft) and
TBR (tree bisection and reconnection) have been proposed in \cite{key-11} 
for measuring the distance between the two phylogenetic trees. In
a pioneering paper, Allen and Steel \cite{key-1}  proposed algorithms for
estimating these distances. The hybridization number and the rooted
SPR (rSPR) distance have proven to be very useful tools for estimating
the reticulation events that have occurred. Baroni et al \cite{key-2}. 
showed that the rSPR distance provides a lower bound on the number of
reticulation events. \\ 

Computing hybridization number, rSPR and TBR distances have been shown
to be NP-hard problems. This triggered interest in designing approximation and fixed parameter 
tractable algorithms for these problems. 
Hein et al. \cite{key-8} introduced the idea of a Maximum
Agreement Forest(MAF) as a new tool to determine the distance between
two phylogenies. They proposed a 3-approximation ratio algorithm
exists for computing a MAF for 2 trees and a NP-hardness
proof for the rSPR distance problem computation. 
Allen and Steel \cite{key-1} showed that the TBR distance between two trees is equal to the number of
components in a MAF. They also corrected an oversight in Hein et al's paper \cite{key-8} to show 
that the TBR problem is NP-hard.      
Rodrigues et al. \cite{key-11} reported a 3-approximation algorithm 
for computing a MAF for 2 trees and generalized this to a $(d+1)$-approximation 
algorithm for computing a MAF for 2 trees with degree at most $d$.
Bonet et al \cite{key-15} pointed out that 3-approximation algorithms claimed by \cite{key-11} and \cite{key-8} 
are in error and proposed a 5-approximation algorithm for the rSPR problem.       
Bordewich and Semple \cite{key-3} showed that the rSPR distance between two rooted trees
is equal to the number of components in a MAF of these trees, and used this result to show the 
computing the rSPR distance is NP-complete.  
In a subsequent paper, Bordewich et al \cite{key-4} proposed a 
3-approximation algorithm for computing the rSPR distance between two trees and a 
fixed-parameter tractable algorithm of time-complexity in $O(4^kk^4+ n^3)$. 
Baroni et al \cite{key-2} introduced the concept of Maximum Acyclic Agreement Forest(MAAF) and
showed that the hybridization number of two trees is one less than the number of components in a MAAF. 
Chataigner \cite{key-5} obtained an 8-approximation ratio algorithm for computing the MAF on \textit{k}
($\geq$2) trees. \\ 

\begin{figure}
\centering
\includegraphics[scale=0.4]{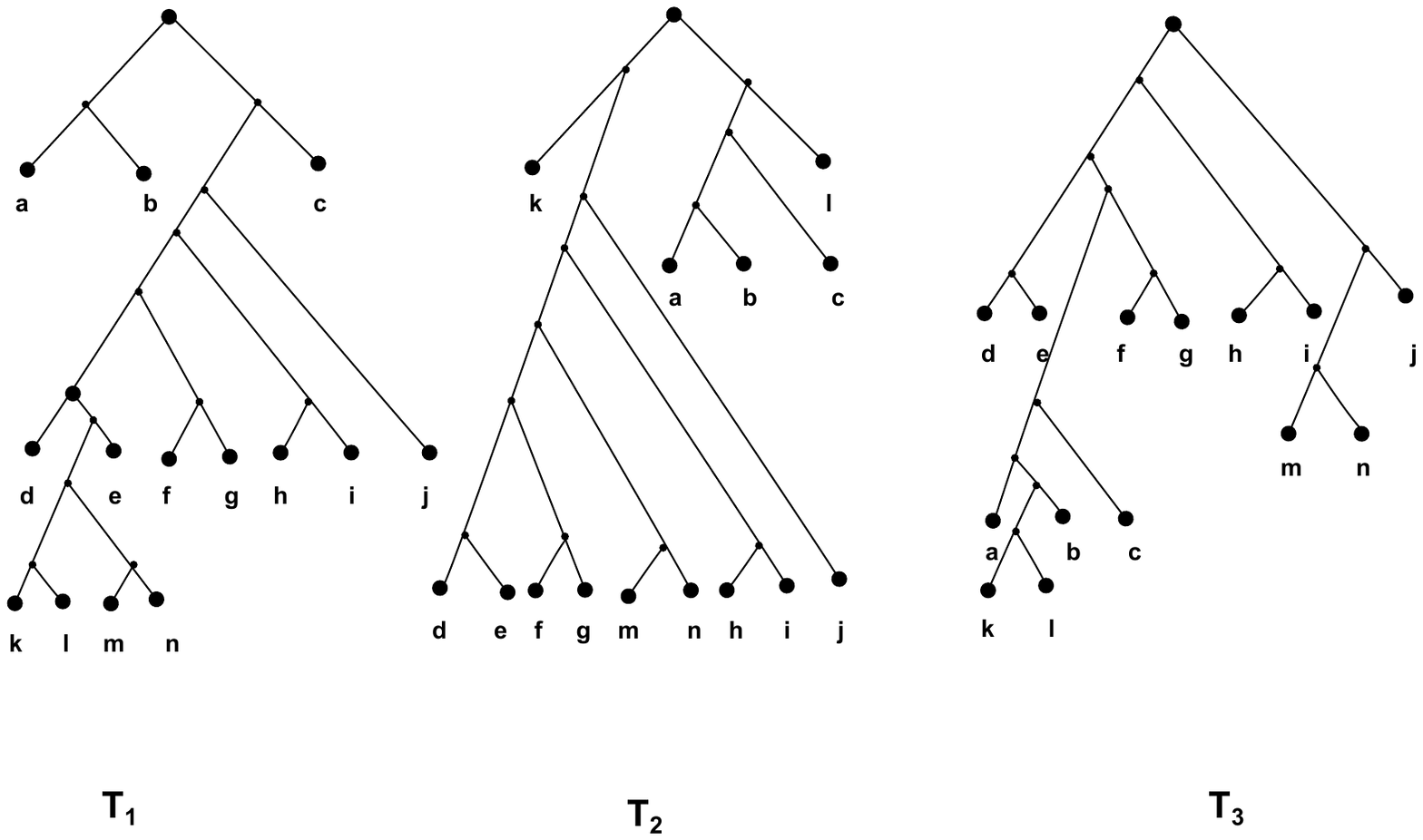}
\caption{\emph{Phylogenetic Trees}}
\label{fig:PhylogenyTrees}
\end{figure}

In this paper, we propose a 3-approximation algorithm for computing  
a MAF on $k (\geq 2)$ trees by a simple extension of Bordewich et al's 3-approximation algorithm 
for 2 trees \cite{key-4}. We also propose a new 3-approximation algorithm for computing a MAAF on $k (\geq 2)$ trees.

\begin{figure}
\centering
\includegraphics[scale=0.2]{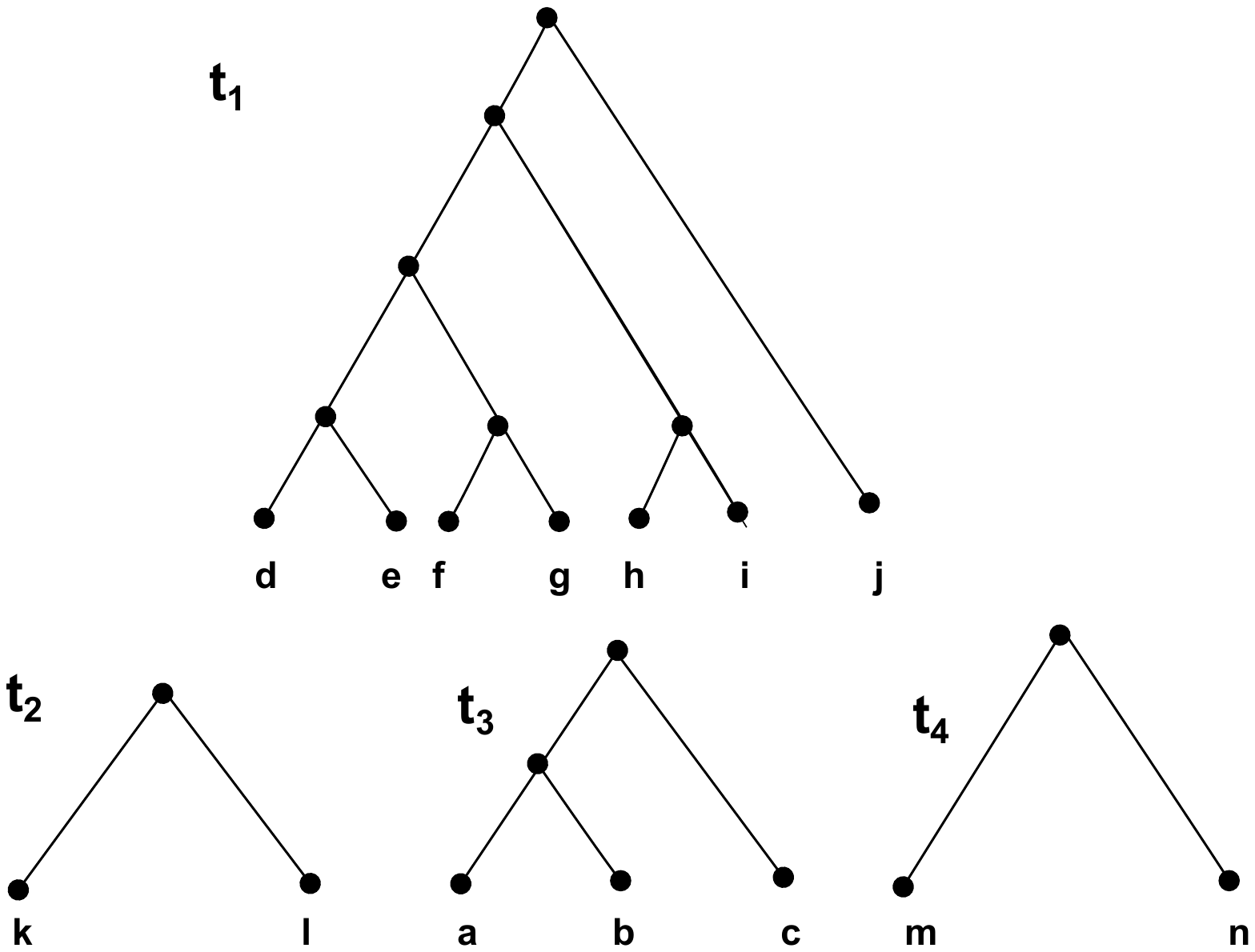}
\caption{\emph{Maximum Agreement Forest of the trees in Fig.1}}
\label{fig:MAF} 
\end{figure}

\section{Preliminaries}
In this section, following {[}1,2,4{]}, we introduce the terminology and notation used in this paper. 

\subsection{Hybridization number}

If \textit{v} is a vertex of a directed graph (digraph, for short) \textit{D}, we
denote the in-degree of $v$ by \textit{d$^{\text{-}}$(v)} and its out-degree by 
and \textit{d$^{\text{+}}$(v)}. A hybrid phylogenetic network, \textit{H}, on an extant set of species \textit{X}
consists of: 

\medskip{}

(i) a \textit{rooted} acyclic digraph \textit{D} in which the root has out-degree
at least two and, for all vertices\textit{ v} with \textit{d$^{\text{+}}$(v)}
= 1, we have \textit{d$^{-}$(v)} $\geq$ 2, and \\  

(ii) the set of vertices of \textit{D} with out-degree zero is precisely \textit{X}. \\ 


\textit{X} is called the \textit{label set} of \textit{H} and is also denoted by \textit{L(H)}. Vertices with 
in-degree at least two are called \textit{hybrid vertices}. 
These vertices represent an exchange
of genetic information between hypothetical ancestors. For a hybrid phylogeny
\textit{H} on \textit{X} with root $\rho$, its \emph{hybridization number}, \textit{h(H)}, is:

\begin{center}
$h(H) = \sum_{v \neq \rho}(d^-(v) - 1)$
\end{center}

\medskip{}

A rooted binary phylogenetic $X$-tree is a special type of hybrid phylogeny
in which the root has degree two and all other interior vertices have degree 
three, while $X$ is its leaf-set. For two rooted binary phylogenetic \textit{X}-trees
\textit{T} and $T'$, we define: 

\begin{center}
\textit{h(T,$T'$) = min \{h(H) : H is a hybrid on X that displays T and $T'$\}}
\end{center}

\subsection{Agreement Forest}

Let $T$ and $T'$ be two rooted binary phylogenetic trees.
We denote the set of leaf labels of $T$ by $L(T)$, and the
set of its edges by $E(T)$. The extent of similarity between the 
two trees can be quantified by computing an \emph{agreement forest}. 
This useful notion, introduced by Hein et al \cite{key-8}, is formally 
defined as follows: \\   

 
An \emph{agreement forest} $F$ for
$T$ and $T'$ is a collection of rooted binary phylogenetic
trees $t_1$, $t_2$,..., $t_n$ such that:

\begin{itemize}
\item for each $i$, $L(t_i) \subseteq L(T)$ and  $\bigcup L(t_i) = L(T)$;

\item for each $t_i$, $S_i$ is the minimal subtree connecting the nodes of $L(t_i)$ in $T$; it  
is identical with $t_i$ when nodes of degree 2 in $S_i$ are contracted;

\item for $i \neq j$, $S_i$ and $S_j$ are node disjoint.
\end{itemize}

The size of an agreement forest is the number of trees (or components) in the forest. 
An agreement forest with the \textit{smallest} number of components is a \emph{Maximum Agreement Forest} (MAF)
for $T$ and $T'$. \\ 

An agreement forest is obtained by cutting the same number of edges from both
\textit{T} and $T'$, followed by contraction of degree 2 nodes from the residual trees. 
The deleted edges are those which do not agree in $T$ and $T'$, suggesting that they represent different 
paths of genetic inheritance i.e. hybridization events (see Figure~\ref{fig:PhylogenyTrees} and Figure~\ref{fig:MAF}). \\  

The notion of a Maximum Acyclic Agreement Forest (MAAF) was introduced in [1] 
to exclude agreement forests in which any vertex in the associated hybrid phylogeny inherits genetic 
information from its own descendants. \\ 

Let $F = \{t_1, t_2, \ldots, t_n \}$ be an agreement forest for $T$ and $T'$.
Let $G_F$ be the directed graph whose vertex set
is $F$ and there is an edge from $t_i$ to $t_j$ if $i \neq j$ and 

\begin{itemize}
\item either the root of $T(L(t_i))$ is an ancestor of the root of $T(L(t_j))$, 
\item or the root of $T'(L(t_i))$ is an ancestor of the root of $T'(L(t_j))$.
\end{itemize}

Note that since $F$ is an agreement forest, 
$T(L(t_i))$ and $T(L(t_j))$ have different roots; the
same is true of $T'(L(t_i))$ and $T'(L(t_i))$. 
We say that $F$ is an \emph{acyclic agreement forest} if $G_F$ is acyclic. 
$F$ is a \emph{Maximum Acyclic Agreement Forest} (MAAF) if
it has the smallest number of components.
In short, an agreement forest is a MAAF if it is a MAF and and  $G_F$ is acyclic. \\ 

The importance of a MAAF stems from the following theorem, proved in [2].

\begin{theorem}
The hybridization number of T and $T'$ is equal to the size
of a MAAF for T and $T'$ minus one.
\end{theorem}

\subsection{Rooted Subtree Prune and Regraft(rSPR)}

Due to reticulation, two phylogenies with the same set of species exhibit 
inconsistencies in their parent-child relationships. 
One approach to quantifying this is to compute the rooted subtree prune and regraft (rSPR) distance between 
two phylogenies [4]. This can be done by a series rSPR operations as explained below. \\ 

\begin{figure}
\centering
\includegraphics[trim=0mm 90mm 0mm 0mm, clip, width = 0.5\columnwidth]{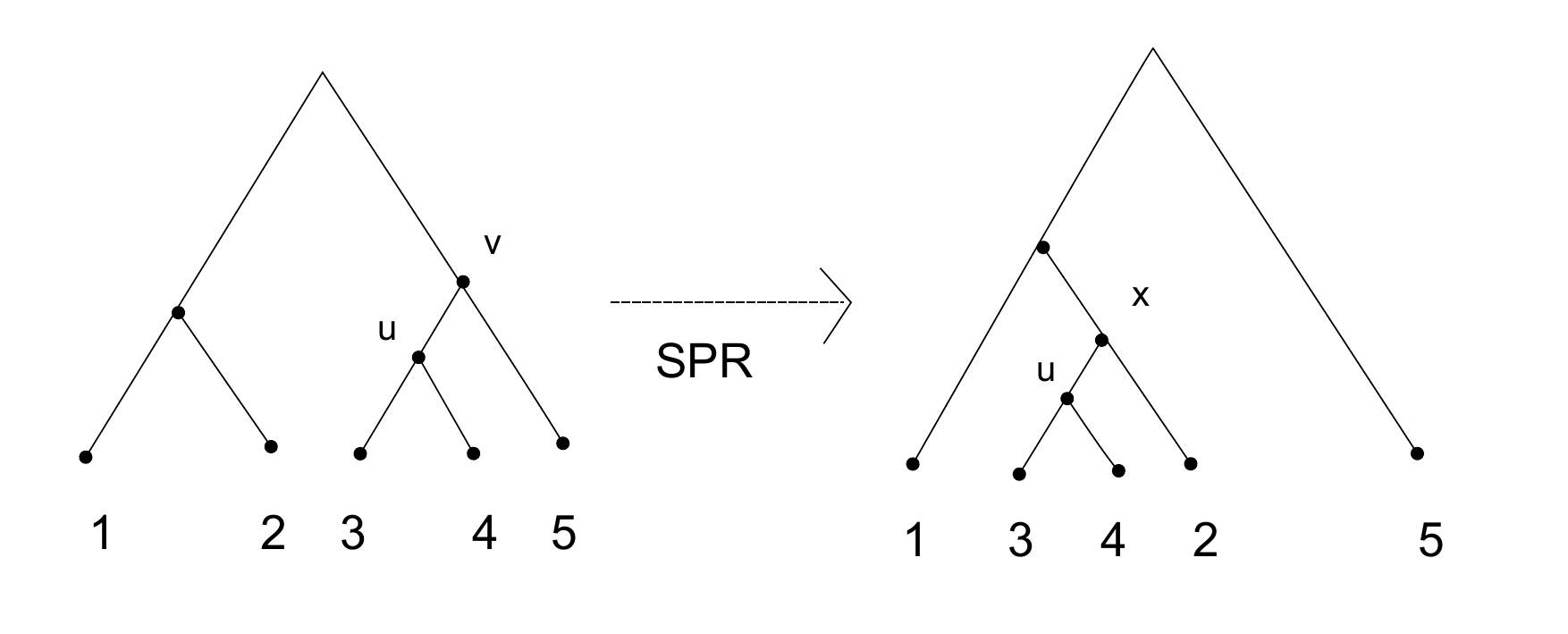}
\caption{\emph{An rSPR operation}}
\label{fig:spr} 
\end{figure}

Figure~\ref{fig:spr} illustrates a typical rSPR operation in which the subtree $T_u$ rooted at $u$ is pruned and grafted 
to another part of the tree by connecting $u$ to a newly-created vertex $x$ (on some pre-existing edge).  
The vertex $v$, now of degree 2, is removed. For more details on rSPR operations see [3].      
rSPR$(T,T')$ measures the minimum number of rSPR operations required to transform $T$ into $T'$ and is equal to the size of a MAF minus 1 (shown in [3]). 

\subsection{Partial order and Incompatible triples}

We define a partial order ($<$) on the edges and vertices (collectively called \emph{elements}) 
of a forest $F$, derived from a phylogenetic tree $T$.
For two distinct elements $x$ and $y$ that belong to the same component of $F$, we have $x < y$ if 
$y$ lies on the path from $x$ to the root of this component.

\begin{figure}[b!]
\centering
\includegraphics[scale=0.3]{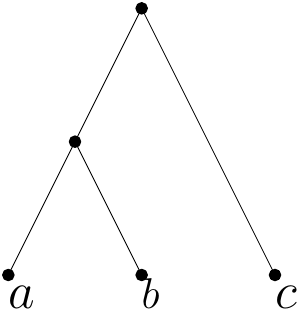}
\caption{\emph{A triple ab$|$c}}
\label{fig:triple} 
\end{figure}

A triple (see Figure~\ref{fig:triple}) is a rooted binary phylogenetic tree that has 3 leaves.
A triple with leaf set $\{a,b,c\}$ is denoted by $ab|c$
if the path from $c$ to the root and the path from $a$ to $b$ are vertex-disjoint. \\ 


Let $\{a,b,c\}$ be a common leaf-set of both $T$ and $T'$.
The triple $ab|c$ is an incompatible triple of $T$ with respect to $T'$ if $ab|c$ is a 
triple of $T$ only. \\  

The partial order defined on elements of $F$ can be extended  
to its incompatible triples. Let $ab|c$ be an incompatible triple
of $T$. If $r_{abc}$ represent the most recent common ancestor of $a$ and $c$ in $T$
and $r_{ab}$ the most recent common ancestor of $a$ and $b$ in $T$, we say $ab|c < xy|z$ if:  

\begin{itemize}
\item either $r_{xyz}$ lies on the path from $r_{abc}$ to the root of $T$ 
\item or if $r_{abc}$ and $r_{xyz}$ are equal, $r_{xy}$ is on the path from $r_{ab}$ to the root of $T$
\end{itemize}

An incompatible triple is \emph{minimal} if it is minimal with respect to this partial order.

\subsection{Inseparable Components}

Let $F$ be the forest obtained from the two rooted binary phylogenetic
trees $T$ and $T'$ after all the incompatible triples have been taken care of. 
If two components $t_x$ and $t_y$ of $F$ share a common element in $T'$,
then $t_x$ and $t_y$ are said to be \textit{inseparable} with respect to $T'$.

\section{Approximation Algorithms} 

Below, by \ptrees~we shall mean rooted \ptrees. 
In this section, we discuss algorithms for approximating a MAF as well as a MAAF for $k$ \ptrees.
We refer to the 3-approximation algorithm for computing an approximate MAF for 2 \ptrees~by Bordewich et al \cite{key-4} 
as Bordewich's algorithm. In the next subsection, we briefly review this algorithm and show how it can be simply 
extended to $k$ \ptrees. In the following subsection, we show how to strengthen one of the key results in \cite{key-4} 
to obtain a 2-approximation algorithm for $k~(\geq 2)$ \ptrees, a substantial improvement over the 8-approximation 
algorithm of Chataigner \cite{key-5}. In the next subsection, we discuss a 2-approximation 
algorithm for computing an approximate MAAF for $k~(\geq 2)$  \ptrees. 

\subsection{Extending Bordewich et al's 3-approximation algorithm to $k$ trees}

The following lemma, proved in [4], plays a central role. Consistent with the phylogeny literature,  
we use the symbols $+$ and $-$ for set-union and set-difference respectively.

\begin{lemma}
$F$ is a forest of an $X$-tree $T$, while
$e$ and $f$ are two edges in the same component of F
such that $f \in E$ and $e \notin E$, where $E$ is a subset of edges of $F$. Let $v_f$
be the end-vertex of $f$ closest to $e$, and $v_e$ an end-vertex
of $e$. If \\   

(i) $v_f \sim v_e$ in F - E and \\  

ii) $v_f \nsim x$ in $F-(E+e)$ for all $x$ in the leaf-set $X$, \\  

then $F-E$ and $F-(E-f+e)$ yield isomorphic forests.
\end{lemma}

\begin{figure}[h!]
\centering
\includegraphics[scale=0.3]{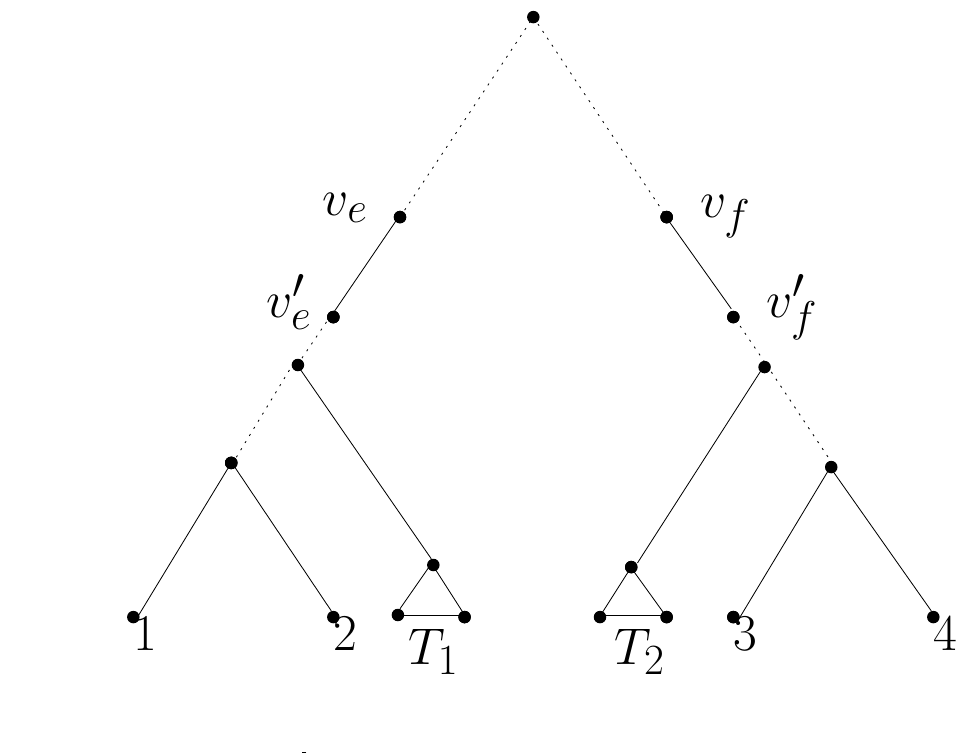}
\caption{\emph{Pictorial illustration of Lemma 1}}
\label{fig:central}
\end{figure}

The essential conclusion of the above theorem is that the edges
$e$ and $f$ are connected by a linear path (see Figure~\ref{fig:central}) in $F - E$. Thus we get 
isomorphic forests by substituting $e$ for $f$ in $E$. \\

The approximation algorithm we present is a simple extension  
of the 3-approximation algorithm in [4] for $k=2$.  
We initialize the agreement forest $F$ to $T_1$. Next, we determine in turn the 
incompatible triples of each $T_i$, $i = 2, 3, \ldots, k$ with respect to $F$ and deletes edges
from $F$ to eliminate the incompatibilities. The incompatible triples are processed with respect to their 
partial order, picking a minimal one from those that remain. \\        

For a minimum incompatible triple $ab|c$ in $F$ with respect to any $T_i~(i \geq 2),$ 
the edges that are candidates for removal are determined as follows (see Figure~\ref{pic:fig5}; this figure is 
based on a similar figure in [4]).   
Let $r_{abc}$ be the most recent common ancestor of $a$ and $c$ in
$T$ and $r_{ab}$ the most recent common ancestor of $a$ and $b$ in $T$. The child
edge of $r_{ab}$ leading to $a$ is denoted by $e_a$ and the child edge of $r_{ab}$
leading to $b$ is denoted by $e_{b}$. We label $e_{r}$ the child edge of $r_{abc}$
leading to $r_{ab}$. Finally, we label by $e_{c}$ the first edge on the path from $r_{abc}$ to $c$
such that for elements $c'$ in the leaf-set of $T-c$ below $e_c$,
there exists triples of the form $cc'|a$ and $cc'|b$ in both $T_i~(i \geq 2)$ and this component of $F$. \\ 
 
\begin{figure}[htbp]
\centering
\includegraphics[scale=0.4]{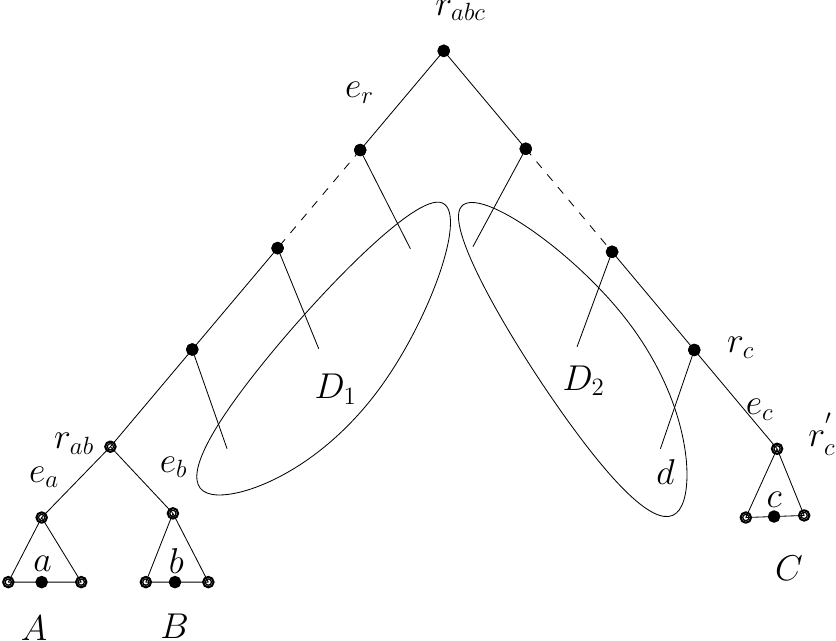}
\caption{\emph{Minimum Incompatible triple} $ab|c$}
\label{pic:fig6}
\end{figure}

After all incompatibilities are resolved, the algorithm determines inseparable components $t_x$ and $t_y$ 
of $F$ vis-a-vis the $T_i$'s ($i \geq 2$) and deletes appropriate edges to eliminate the overlaps.         
For a pair of inseparable components $t_x$ and $t_y$ in any $T_i$, $i = 2, 3, \ldots, k$ with respect to 
$F$, let $v_{xy}$ denote a minimal common vertex of $t_x + t_y$ in $T_i$
with respect to the partial order on the vertices in $T_i$.
Further, $e_x$ denotes the minimal edge in $F$ whose descendants in the leaf-set are also descendants of
$v_{xy}$ in $t_x$. Similarly, $e_y$ denotes the minimal edge in $F$ whose set of
descendants in the leaf-set are also the descendants of $v_{xy}$
in $t_y$ (see Figure~\ref{pic:fig6}).

\begin{figure}[t!]
\centering
\includegraphics[scale=0.4]{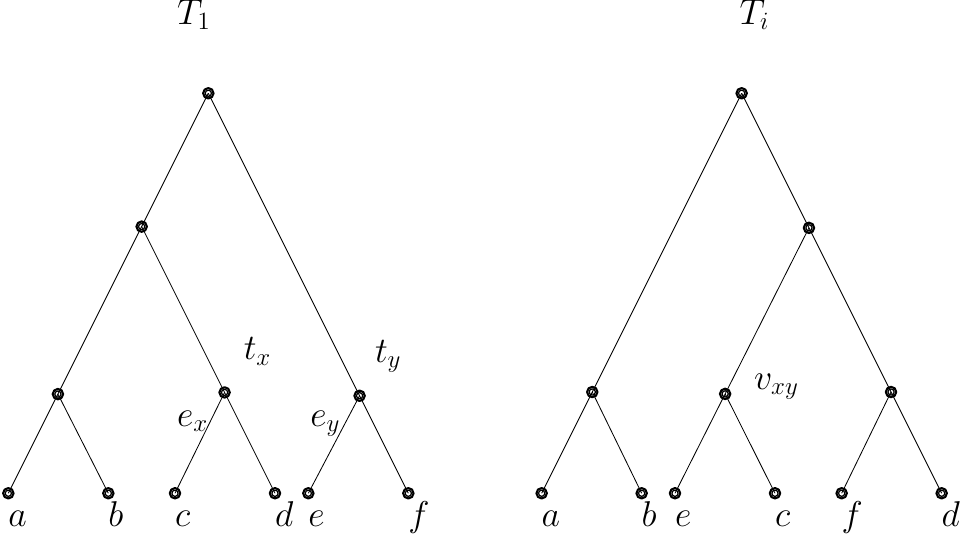}
\caption{\emph{Components $t_x$ and $t_y$ of $ F (= T_1)$ overlap in $T_i$, for some $i \geq 2$}}
\label{pic:fig7}
\end{figure}


\vspace{0.5cm}
\hrule
\vspace{0.5cm}

{\bf Algorithm} \emph{MAF-Approx($T_1,T_2,...,T_k$)} \\

1. $F \leftarrow T_1$;

2. for $i = 2$ to $k$ do


\qquad{}2.1. while there exists an incompatible triple in $F$ with respect to $T_i$ do


\qquad{}\qquad{}2.1.1. consider the minimal incompatible triple
$ab|c$ in $F$ with respect to $T_i$ 

\qquad{}\qquad{}2.1.2. $E \leftarrow \{e_a, e_c, e_r\}$ in $ab|c$

\qquad{}\qquad{}2.1.3. $F \leftarrow F - E$

\enskip{}\enskip{}\qquad{}enddo;

\enskip{}\enskip{}enddo;

3. for $i = 2$ to $k$ do


\qquad{}3.1. while there exists a pair of inseparable components in any $T_i$
($i \geq 2$) with respect to $F$ do


\qquad{}\qquad{}3.1.1. consider inseparable components $t_x$
and $t_y$ in $T_i$ with respect to $F$

\qquad{}\qquad{}3.1.2. $E \leftarrow \{e_x, e_y \}$ in $t_x$ and $t_y$

\qquad{}\qquad{}3.1.3. $F \leftarrow F - E$

\enskip{}\enskip{}\qquad{}enddo;

\enskip{}\enskip{}enddo;

4. return \textit{F};

\vspace{0.5cm}
\hrule
\vspace{0.5cm}

\begin{lemma}
Let $T_1, T_2, \ldots, T_k$ be $k$ rooted binary phylogenetic
trees and F a forest of $T_1$. \\  

(i) If $ab|c$ is a minimal incompatible triple of F  
with respect to some of the $T_i$'s for ($i \geq 2$), then 

\begin{center}
$e(F-\{e^i_{a}, e^i_{c}, e^i_{r}\}, T_2,T_3,...,T_i,...,T_k) \leq  e(F, T_2, T_3,...,T_i,...,T_k) - 1$,
\end{center}
 
where $e(F, T_2, T_3,...,T_k)$ denotes the size of a minimum
set $E$ of edges of $F$ such that $F - E$ forms an agreement forest of $k$
trees and $e^i_{a}, e^i_{c}, e^i_{r}$ are the edges deleted from $F$ to resolve the incompatibility 
due to the triple $ab|c$. \\ 

(ii) If there is no incompatible triple of $F$ with respect to
any other tree, but there exist two components $t_x$ and
$t_y$ of $F$ that overlap in some tree $T_i$ ($i \geq 2$), then
for some $j \in \{x, y\}$ 

\begin{center}
$e(F - e^i_{j}, T_2, T_3,...,T_i,...,T_k) = e(F, T_2,T_3,...,T_i,...,T_k)$ - 1.
\end{center}

\end{lemma}

{\bf Proof:} The proof in [4] for the case when $k = 2$ goes through with some minor changes. Let's 
see this for (i). \\  

We have $|E| = e(F, T_2, T_3, \ldots, T_k)$, where $E$ is a minimum set of
edges such that $F-E$ yields a maximum agreement forest of $F$ and all $T_i$, $i \geq 2$. It has been shown in [4]
that there exists an $f \in E$ such that $F - (E - f + \{e_a, e_c, e_r \})$ is isomorphic to a subforest of $F - E$. 
This implies that  $F - (E - f + \{e_a, e_c, e_r \})$ yields an agreement forest of $F - \{e_a, e_c, e_r \}$ and all $T_i,~i \geq 2$. 
So, $e(F - \{e_a, e_c, e_r \}) \leq |E-f| = e(F, T_2, T_3, \ldots, T_k) - 1$. \\    

We can likewise extend the proof of (ii) for the case when $k = 2$ to this case. $\hfill \Box$ \\  

Assume there are $\alpha_1$ iterations of the 1st while loop and $\alpha_2$ iterations
of the 2nd while loop in processing all $k-1$ trees, $T_2, T_3, \ldots, T_k$. 
Set $\alpha = \alpha_1 + \alpha_2$. Then we have the following claim. 

\begin{claim}
\textit{$\alpha$ $\leq$ e(T$_{\text{1}}$,T$_{\text{2}}$,T$_{\text{3}}$,...,T$_{\text{k}}$)
$\leq$ 3$\alpha$}.  
\end{claim}

{\bf Proof:} Let $F_i$ be the forest obtained after $i$ iterations of the above algorithm. The following cases arise. \\  

{\bf Case 1:} $i \leq \alpha_1 $ (still in 1st while loop) 

From Lemma 2, we have:  

\begin{center}
\textit{e(F$_{i}$,T$_{\text{2}}$,T$_{\text{3}}$,...,T$_{\text{k}}$)
$\leq$ e(F$_{i-1}$,T$_{\text{2}}$,T$_{\text{3}}$,...,T$_{\text{k}}$)
}- 1
\end{center}

Hence by successive applications of the above inequality, we have: 

\begin{center}
 \textit{e(F$_{i}$,T$_{\text{2}}$,T$_{\text{3}}$,...,T$_{\text{k}}$)
+ i $\leq$ e(F$_{0}$,T$_{\text{2}}$,T$_{\text{3}}$,...,T$_{\text{k}}$)} 
\end{center}

Since $F_0 = T_1$, we can rewrite the above inequality as:

\begin{center}
\textit{e(F$_{i}$,T$_{\text{2}}$,T$_{\text{3}}$,...,T$_{\text{k}}$)
+ i $\leq$ e(T$_{1}$,T$_{\text{2}}$,T$_{\text{3}}$,...,T$_{\text{k}}$)}
\end{center}

Moreover, since \textit{F$_{i}$} has 3 fewer edges than \textit{F$_{i-1}$} we have:

\begin{center}
\textit{e(F$_{i-1}$,T$_{\text{2}}$,T$_{\text{3}}$,...,T$_{\text{k}}$)
$\leq$ e(F$_{i}$,T$_{\text{2}}$,T$_{\text{3}}$,...,T$_{\text{k}}$)
}+ 3
\end{center}

Applying the above $i$ times, we get 

\begin{center}
\textit{e(F$_{0}$,T$_{\text{2}}$,T$_{\text{3}}$,...,T$_{\text{k}}$)
$\leq$ e(F$_{i}$,T$_{\text{2}}$,T$_{\text{3}}$,...,T$_{\text{k}}$)
+ 3i}
\end{center}

and since $F_0 = T_1$,  

\begin{center}
\textit{e(T$_{1}$,T$_{\text{2}}$,T$_{\text{3}}$,...,T$_{\text{k}}$)
$\leq$ e(F$_{i}$,T$_{\text{2}}$,T$_{\text{3}}$,...,T$_{\text{k}}$)
+ 3i}
\end{center}

{\bf Case 2:} \textit{i} $>$$\alpha$$_{\text{1}}$, (in 2nd while loop)

Again, from the second part of Lemma 2,   

\begin{center}
\textit{e(F$_{i}$,T$_{\text{2}}$,T$_{\text{3}}$,...,T$_{\text{k}}$)
$\leq$ e(F$_{i-1}$,T$_{\text{2}}$,T$_{\text{3}}$,...,T$_{\text{k}}$)}
- 1
\end{center} 

Thus by $i$ applications of the above inequality we have:

\begin{center}
\textit{e(F$_{i}$,T$_{\text{2}}$,T$_{\text{3}}$,...,T$_{\text{k}}$)
+ i $\leq$ e(F$_{0}$,T$_{\text{2}}$,T$_{\text{3}}$,...,T$_{\text{k}}$) }
\end{center} 

Since $F_0 = T_1$, we can rewrite the above inequality as:

\begin{center}
\textit{e(F$_{i}$,T$_{\text{2}}$,T$_{\text{3}}$,...,T$_{\text{k}}$)
+ i $\leq$ e(T$_{1}$,T$_{\text{2}}$,T$_{\text{3}}$,...,T$_{\text{k}}$)}
\end{center} 

Again, since \textit{F$_{i}$} has 2 fewer edges than \textit{F$_{i-1}$} we have:

\begin{center}
\textit{e(F$_{i-1}$,T$_{\text{2}}$,T$_{\text{3}}$,...,T$_{\text{k}}$)
$\leq$ e(F$_{i}$,T$_{\text{2}}$,T$_{\text{3}}$,...,T$_{\text{k}}$)
}+ 2
\end{center} 

Thus, 

\begin{center}
\textit{e(T$_{1}$,T$_{\text{2}}$,T$_{\text{3}}$,...,T$_{\text{k}}$)
$\leq$ e(F$_{i}$,T$_{\text{2}}$,T$_{\text{3}}$,...,T$_{\text{k}}$)
+ 3$\alpha$$_{\text{1}}$ + 2(i - $\alpha$$_{\text{1}}$)}
\end{center}

Thus on termination of both loops, we have:

\begin{center}
\textit{e(F$_\alpha$,T$_{\text{2}}$,T$_{\text{3}}$,...,T$_{\text{k}}$)
+ $\alpha$$_{\text{1}}$ + $\alpha$$_{\text{2}}$ $\leq$ e(T$_{1}$,
T$_{\text{2}}$,T$_{\text{3}}$,...,T$_{\text{k}}$) $\leq$ e(F$_\alpha$,
T$_{\text{2}}$,T$_{\text{3}}$,...,T$_{\text{k}}$) + 3$\alpha$$_{\text{1}}$
+ 2$\alpha$$_{\text{2}}$}
\end{center} 

As no further edge-cut is necessary when an agreement forest is generated, 
\textit{e(F$_\alpha$,T$_{\text{2}}$,T$_{\text{3}}$,...,T$_{\text{k}}$)}
= 0. Thus: 

\begin{center}
\textit{$\alpha$$_{\text{1}}$ + $\alpha$$_{\text{2}}$ $\leq$
e(T$_{1}$,T$_{\text{2}}$,T$_{\text{3}}$,...,T$_{\text{k}}$) $\leq$
3$\alpha$$_{\text{1}}$ + 2$\alpha$$_{\text{2}}$}
\end{center}

A fortiori, we have:

\begin{center}
\textit{$\alpha$$\leq$ e(T$_{1}$,T$_{\text{2}}$,T$_{\text{3}}$,...,T$_{\text{k}}$)
$\leq$3$\alpha$$_{\text{1}}$ + 3$\alpha$$_{\text{2}}$}, 
\end{center}

which simplifies to:

\begin{center} 
\textit{$\alpha$$\leq$ e(T$_{1}$,T$_{\text{2}}$,T$_{\text{3}}$,...,T$_{\text{k}}$)
$\leq$3$\alpha$}
\end{center} 

This proves the claim. \hfill $\Box$ \\

Since the number of edges removed is 
$3\alpha_1 + 2 \alpha_2 \leq 3 \alpha \leq 3 e(T_1,T_2,T_3,...,T_k)$, 
our algorithm has an approximation ratio of 3. Summarizing the above results, we have:

\begin{theorem}
Algorithm  MAF-approx has an approximation ratio of 3 and time-complexity in $O(kn^5)$.   
\end{theorem}

\section{A 3-approximation algorithm for computing an approximate MAAF}

\begin{figure}[h!]
\centering
\includegraphics[scale=0.3]{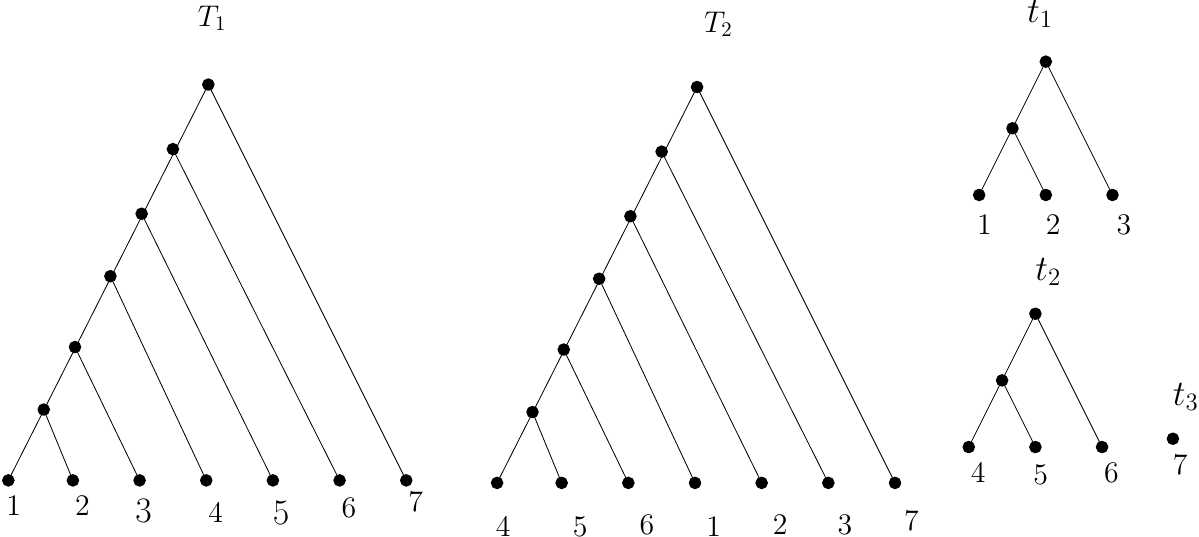}
\caption{\emph{Example of a Maximum Agreement Forest of trees $T_1$ and $T_2$}}
\label{fig8}
\end{figure}

The roots of two components (trees) in a MAF produced by the algorithm of the previous section may have 
an ancestor-descendant relationship in one tree, and the opposite in another, as in Figure~\ref{fig8}. 

If there are no cycles in the approximate MAF, $F$, produced by the algorithm \emph{MAF-Approx} 
of the previous section then we have an approximate MAAF, $F_A$, as well. 
Otherwise, as in      
Whidden and Zeh [13], we incorporate a preprocesing step to detect cycles that exist between roots of the 
trees in the approximate MAF and eliminate these cycles. This consists of assigning to each node of $T_i, i=1,2, \ldots,k$, 
a preorder visit number and the interval of preorder numbers of its descendants. 

We maintain the roots of all the trees in the MAF in 2 sets: $R_p$ and $R_{up}$. The former consists of the roots of all 
trees with no cycle between any pair; the roots of the latter are yet to be processed. We choose a root from $R_{up}$ and determine 
if it has a cycle with any root in $R_{p}$. This is done by mapping these roots to the corresponding nodes of a pair of trees 
$T_i$ and $T_j$. 
The preorder intervals associated with these nodes can be used to check the existence of a cycle between these roots (for more details see
\cite{key-13}).     
If a cycle does not exist, we add it to the 
set $R_{p}$. Otherwise, let $t_i$ and $t_j$ be 2 trees in $F$ whose respective roots $r$ and $r'$ form a cycle. 
We call such a tree-pair \emph{infeasible}.
To obtain \textit{F$_{\text{A}}$}, we delete one of the edges $e_r$ incident on $r$, as well as one of the edges $e_{r'}$ 
incident on $r'$ to remove this cycle. We continue doing this till the set $R_{up}$ becomes empty. 
The following lemma underlies the above choice of the edges we cut.    

\begin{lemma}
If $e(F, T_2, \dots, T_k)$ is the minimum number of edges that must be removed from $F$ to obtain a MAAF, then 
$e(F - \{e_x\}, T_2, \ldots, T_k) = e(F, T_2, \dots, T_k) - 1$, where $x \in \{r,r'\}$; moreover, 
$e(F - \{e_r, e_{r'}\}, T_2, \ldots, T_k) \leq  e(F, T_2, \dots, T_k) - 1$
\end{lemma}

\textbf{Proof:} We prove that there  exists a set $E$ of $e(F, T_2, \dots, T_k)$ edges of $F$ such that $F-E$ yields a 
MAAF of $T_1, T_2, \ldots, T_k$ and $E \cap \{e_r, e_{r'}\} \neq \emptyset$. Otherwise, let $E$ be a set of edges 
that produces a MAAF but  $E \cap \{e_r, e_{r'}\} = \emptyset$. \\ 

Let $t_i(r_1)$ and $t_i(r_2)$ be the left and right subtrees of $t_i(r)$, and $t_j(r'_{1})$ and $t_j(r'_{2})$ the 
left and right subtrees of $t_j(r')$.  
There exists $i$ such that $a' \nsim_{F-E} r$ for all $a' \in X_{T(r_i)}$ or  
$b' \nsim_{F-E} r'$ for all $b' \in X_{T(r_i)}$. \\ 

Otherwise, there exists $a_1 \in t_i(r_1)$ and $a_2 \in t_i(r_2)$ such that $a_1 \sim a_2$ and 
$b_1 \in t_j(r'_1)$ and $b_2 \in t_j(r'_2)$ such that $b_1 \sim b_2$. This implies that the 
cycle involving the roots $r$ and $r'$ (of $t_i$ and $t_j$ respectively) have not been removed. \\ 

Assume that $a' \nsim_{F-E} r$ for all $a' \in T(r_1)$. Choosing an edge $f$ closest to $r$ on the 
path from $r$ to a leaf $a'$, by Lemma 1, the forests $F-E$ and $F-(E-f + e_r)$ are isomorphic. The claims
of the lemma follow from this.        
   
$\hfill \Box$ \\

A formal description of the above algorithm is given below. 

\vspace{0.5cm}
\hrule
\vspace{0.5cm}

{\bf Algorithm} \emph{MAAF-Approx($F$)} \\

// $F = \{t_1, t_2, t_3, \ldots, t_m \}$  

1. Set $R_{up} \leftarrow \{ root(t_1), root(t_2), \ldots, root(t_m) \}$ 

2. Set $R_{p}  \leftarrow \emptyset $ 

3. while \{$R_{up} \neq \emptyset \}$  

\hspace{0.3cm} do 

\hspace{0.3cm} 3.1 Pick an $r$ from $R_{up}$

\hspace{0.3cm} 3.2 If ($r$ forms a cycle with an $r'$ in $R_{p}$) then 
 
\hspace{0.7cm} 3.2.1 Delete an edge $e_r$ incident on $r$ and an edge $e_{r'}$ incident on $r'$
  
\hspace{0.7cm} 3.2.2  Add roots of the subtrees of $r$ and $r'$ to $R_{up}$ 
  
\hspace{0.7cm} 3.2.3  Continue 
 
\hspace{0.3cm} 3.3 else $R_{p} \leftarrow R_{p} + r$  

\hspace{0.3cm} od

4. Return the trees whose roots are in $R_p$  

\vspace{0.5cm}
\hrule
\vspace{0.5cm}


Assume there are $\beta$ iterations of  \emph{MAF-Approx} and \emph{MAAF-Approx} in each of which at most 3 edges are removed.  
We claim that: 


\begin{claim}
$\beta \leq e(T_1, T_2, \ldots, T_k) \leq  3\beta $

\end{claim}

{\bf Proof:} 
Let $F_\theta$ be the forest generated after $\theta$ iterations. 
By Lemmas 2 and 3,  $e(F_\theta, T_2, T_3, \ldots, T_k)) \leq e(F_{\theta-1}, T_2, T_3, \ldots, T_k) - 1$.
Hence after $\beta$ iterations, $e(F_{\beta}, T_2, T_3, \ldots, T_k) + \beta \leq  e(T_1, T_2, T_3, \ldots, T_k)$ 
(as $e(F_0, T_2, T_3, \ldots, T_k) = e(T_1, T_2, T_3, \ldots, T_k)$). \\ 

Conversely, the algorithms $MAF-Approx(T_1, T_2,...,T_k)$ as well as $MAAF-Approx(F)$ account for at most 3 
edge-cuts in each iteration. Hence, $e(F_{\theta-1}, T_2, T_3, \ldots, T_k) \leq e(F_\theta, T_2, T_3, \ldots, T_k) + 3$.
This implies $e(T_1, T_2, T_3, \ldots, T_k) \leq e(F_{\beta}, T_2, T_3, \ldots, T_k)+ 3\beta$. \\ 

Now, after $\beta$ iterations an Acyclic-MAF is generated and we
do not require any further edge-cuts. So, $e(F_{\beta}, T_2, T_3, \ldots, T_k) = 0$. \\ 

This proves that $\beta \leq e(T_1, T_2, T_3, \ldots, T_k) \leq 3\beta$ and consequently that our algorithm 
has approximation ratio 3.

\hfill $\Box$ \\

Summarizing the above discussions, we have: 


\begin{theorem}
Algorithm MAAF-approx has an approximation ratio of 2 and time complexity in $O(n^2k^2)$.
\end{theorem}


\section{Conclusion}

In this paper we have proposed approximation algorithms for finding
the Maximum Agreement Forest and the Maximum Acyclic Agreement Forest
on $k$ rooted phylogenetic trees. It is straightforward to extend the fixed-parameter 
tractable algorithm for an exact MAF of 2 trees \cite{key-4} to $k (\geq 2)$ trees. \\

Extending these algorithms to $k$ unrooted trees
would be interesting, as would be to extend the results to trees of degree $d$ ($d \geq 2$). 


\end{document}